\RequirePackage{fix-cm}
\documentclass[twocolumn,epjc3]{svjour3}  
\smartqed  % flush right qed marks, e.g. at end of proof
\RequirePackage{graphicx}
\RequirePackage{mathptmx}      % use Times fonts if available on your TeX system
\RequirePackage{latexsym}
\RequirePackage[numbers,sort&compress]{natbib}
\RequirePackage[colorlinks,citecolor=blue,urlcolor=blue,linkcolor=blue]{hyperref}

\usepackage{multirow}
\usepackage{makecell}
\usepackage[T1]{fontenc}
\usepackage{array}

\usepackage{newtxtext,newtxmath}

\usepackage{url}
\usepackage{hyperref}

\usepackage[usenames,dvipsnames]{xcolor}

\journalname{Eur. Phys. J. C}
\begin{document}

\title{Inferring $S_8(z)$ and $\gamma(z)$ with cosmic growth rate measurements using machine learning}
%\subtitle{Do you have a subtitle?\\ If so, write it here}

%\titlerunning{Short form of title}        % if too long for running head

\author{Felipe Avila\thanksref{e1,addr1}
        \and
        Armando Bernui\thanksref{addr1} 
        \and
        Alexander Bonilla\thanksref{addr2}
        \and
        Rafael C. Nunes\thanksref{addr3, addr4}
}

%\thankstext{t1}{Grants or other notes
%about the article that should go on the front page should be
%placed here. General acknowledgments should be placed at the end of the article.
\thankstext{e1}{e-mail: felipeavila@on.br}

%\authorrunning{Short form of author list} % if too long for running head

\institute{Observat\'orio Nacional, Rua General Jos\'e Cristino 77, 
	S\~ao Crist\'ov\~ao, 20921-400 Rio de Janeiro, RJ, Brazil \label{addr1}
           \and
           Departamento de F\'isica, Universidade Federal de Juiz de Fora, 36036-330, Juiz de Fora, MG, Brazil. \label{addr2}
           \and
           Instituto de F\'{i}sica, Universidade Federal do Rio Grande do Sul, 91501-970 Porto Alegre RS, Brazil \label{addr3}
           \and
           Instituto Nacional de Pesquisas Espaciais, Av. dos Astronautas 1758, 
	Jardim da Granja, S\~ao Jos\'e dos Campos, SP, Brazil\label{addr4}
}

\date{Received: date / Accepted: date}
% The correct dates will be entered by the editor

\maketitle

\begin{abstract}
Measurements of the cosmological parameter $S_8$ provided by cosmic microwave background and large scale structure data reveal some tension between them, 
suggesting that the clustering features of matter in these early and late cosmological tracers could be different. In this work, we use a supervised learning 
method designed to solve Bayesian approach to regression, known as Gaussian Processes regression, to quantify the cosmic evolution of $S_8$ up to 
$z \sim 1.5$. 
For this, we propose a novel approach to find firstly the evolution of the function $\sigma_8(z)$, then we find the function $S_8(z)$. 
As a sub-product we obtain a minimal cosmological model-dependent $\sigma_8(z=0)$ and $S_8(z=0)$ estimates. 
We select independent data measurements of the growth rate $f(z)$ and of $[f\sigma_8](z)$ according to criteria of non-correlated data, 
then we perform the Gaussian reconstruction of these data sets to obtain the cosmic evolution of $\sigma_8(z)$, $S_8(z)$, and the growth index $\gamma(z)$. 
Our statistical analyses show that $S_8(z)$ is compatible with Planck $\Lambda$CDM cosmology; when evaluated at the present time we find 
$\sigma_8(z=0) = 0.766 \pm 0.116$ and $S_8(z=0) = 0.732 \pm 0.115$. 
Applying our methodology to the growth index, we find $\gamma(z=0) = 0.465 \pm 0.140$. 
Moreover, we compare our results with others recently obtained in the literature. 
In none of these functions,  i.e. $\sigma_8(z)$, $S_8(z)$, and $\gamma(z)$, do we find significant deviations from the standard cosmology predictions. 
\end{abstract}

\section{Introduction}

The way how the matter clusters throughout the universe evolution is one of the critical probes 
to judge whether the concordance model $\Lambda$CDM is, in fact, the standard model of 
cosmology. In front of this scenario, accurate measurements of $f(z)$, the growth rate of cosmic 
structures, and of $\sigma_{8}(z)$, the variance of the matter fluctuations at the scale of 
8~Mpc$/h$, are important scientific targets of current and future large astronomical surveys~\citep{Pezzotta17,Aubert20,Bautista21}.

The growth rate, $f$, represents a measure of the matter clustering evolution from 
the primordial density fluctuations to the large-scale structures observed today, 
as such it behaves differently in $\Lambda$CDM-type models, based on the theory of 
general relativity (GR), and in alternative models of cosmology, based on modified gravity theories. 
%(see, e.g.,~\cite{Linder07,Basilakos12,Perenon15,Perenon17,Perenon19,Benisty21a,Benisty21b,Motta21,}). 
On the other hand, $\sigma_{8,0} \equiv \sigma_{8}(z=0)$ can be obtained using the cosmic microwave background 
(CMB) data, where it scales the overall amplitude of the measured angular power 
spectrum\footnote{The observed CMB angular power spectrum amplitude scales nearly 
proportional with the primordial comoving curvature power spectrum amplitude $A_s$, 
but assuming the $\Lambda$CDM model this amplitude constraint can be converted into 
the fluctuation at the present day, usually quantified by the 
$\sigma_{8,0}$ parameter.}~\citep{Planck20}.

The growth rate of cosmic structures is defined as $f(a) \equiv d\ln D(a) / d\ln a$, 
where $D = D(a)$ is the linear growth function, and $a$ is the scale factor in the Robertson-Walker metric, 
based on GR theory. 
A direct measurement of $f$ applying the above relationship to a given data set does not work 
because the cosmological observable is the density contrast and not the growth function $D(a)$~\citep{Avila22}. 
However, it is possible to obtain indirect measurements of $f$ if one can measure the velocity 
scale parameter $\beta \equiv f / b$, and one knows the linear bias $b$ of the cosmological 
tracer used in the measurement of $\beta$~\citep{Bilicki11,Boruah20,Said20,Avila21}. 
Additionally, the most common approach to quantify the clustering evolution of cosmic 
structures is in the form of their product~\footnote{Usually, $f\sigma_{8}$ is termed the 
parametrized growth rate.}, $[f\sigma_{8}](z)$, through the analyses of the Redshift Space 
Distortions (RSD) \citep{Perenon20}, that is, studying the distortions in the two-point correlation function (2PCF) 
caused by the Doppler effect of galaxy peculiar velocities, associated with the gravitational 
growth of inhomogeneities~\citep{Kaiser87} 
\citep[for other applications of the 2PCF in matter clustering analyses see, e.g.,][]{Avila18,Avila19,
Pandey20,Pandey21b,deCarvalho20,deCarvalho21}.

Efforts done in recent years have provided measurements of both quantities: 
$f\sigma_{8}$ and $f$, at various redshifts and through the analyses of a diversity of cosmological tracers, 
including luminous red galaxies, blue galaxies, voids, and quasars. We shall explore these data to find, 
as robust as possible, a measurement of $\sigma_{8}(z)$ and $S_8(z)$, quantities that has been reported to be 
in some tension when comparing the measurements from the last Planck CMB data release~\citep{Planck20} with 
the analyses from several large-scale structure surveys~\citep{Amico20,Philcox_2020,Garcia21,Valentino_2021_S8,perivolaropoulos2021challenges,Huang21,Nunes2021S8}.

The main objective of our analyses is to break the degeneracy in the product function 
$[f\sigma_{8}](z)$ using the cosmic growth rate data $f(z)$, to know the evolution of the functions $\sigma_{8}(z)$ and $S_8(z)$. 
In turn, the knowledge of $\sigma_{8}(z)$ provides its value at $z = 0$, $\sigma_{8,0}$, an interesting outcome of these  analyses 
considering the current $\sigma_{8}$-tension reported in the 
literature~\citep{Valentino_2021_S8,perivolaropoulos2021challenges,Nunes2021S8}. 
Our approach consists of using the Gaussian processes tool to reconstruct the functions $[f\sigma_{8}](z)$ and $f(z)$, 
using for this task two data sets: 
20 measurements of $[f\sigma_{8}](z)$ and 11 measurements of $f(z)$, respectively. 
The reconstructed functions $[f\sigma_{8}]^{\mbox{\sc \,gp}}(z)$ and $f^{\mbox{\sc \,gp}}(z)$ allow us to know 
the function $\sigma_{8}(z)$, as described in the next section.

This work is organized as follows. In section~\ref{sec2} we review the main equations of the linear 
theory of matter perturbations. 
In section~\ref{data} we present the data sets and describe the statistical methodology used in our analyses. 
Section~\ref{results} we report our main results and discussions. 
We draw our concluding remarks in Section~\ref{sec5}. 

%%-----------------------------------------------------------------------
\section{Theory}\label{sec2}

On sub-horizon scales, in the linear regime, and assuming that dark energy does not cluster, the evolution 
equation for the growth function is given by 
\begin{eqnarray}
\frac{df(a)}{d\ln a} + f^2 + \left ( 2 + \frac{1}{2} \frac{d \ln H(a)^2}{d \ln a} \right ) f - \frac{3}{2} \Omega_m(a) = 0\,,
\end{eqnarray}
where $\Omega_m(a) \equiv \Omega_{m,0}\,a^{-3}H_0^2/H(a)^2$, with $\Omega_{m,0} \equiv \Omega_m(z=0)$ 
the matter density parameter today, and $H(a)$ is the Hubble rate as a function of the scale factor, $a$. 
A good approximation for $f(z)$ is given by~\citep{Wang98,Amendola04, Linder05}
\begin{equation}\label{Lindeapprox}
f(z) \simeq  \Omega_{m}^{\gamma}(z) \,,
\end{equation}
where $\gamma$ is termed the growth index.  
For dark energy models within GR theory $\gamma$ is considered a constant with approximate value 
$\gamma \simeq 3(\omega - 1)/(6\omega - 5)$~\citep{Linder07}. 
In the $\Lambda$CDM model, where $\omega = -1$, one has $\gamma = 6/11 \simeq 0.55$. 
However, in alternative cosmological scenarios the growth index can indeed assume distinct functional 
forms beyond the constant value~\citep{Linder07,Batista14}. 
In fact, from equation (\ref{Lindeapprox}) one can define, 
\begin{equation}\label{gammadef}
\gamma(z) \equiv \frac{\ln f(z)}{\ln \Omega_{m}(z)} \,,
\end{equation}
a more general definition for $\gamma$. 
%%-------------------------------------------------------------------------------------

The mass variance of the matter clustering is given by
\begin{equation}\label{sigR}
\sigma^2_R(z) = \frac{1}{2\pi^2}\int_{0}^{\infty}\!P(k, z)W_R^2(k)dk \,,
\end{equation}
where $P(k, z)$ is the matter power spectrum and $W_R(k)$ is the window function with $R$ 
symbolizing a physical scale. 
The matter power spectrum can be written as 
\begin{equation}\label{Pkz}
P(k, z) = \left[\frac{D(z)}{D(z=0)}\right]^2 T^2(k)P(k,z=0) \,,
\end{equation}
where $T^2(k)$ is the transfer function. 
One can write the equation (\ref{sigR}) as
\begin{equation}\label{sigRz}
\sigma_R^2(z) = D^2(z)\sigma_R^2(z=0) \,,
\end{equation}
assuming the normalization $D(z=0) = 1$ for the linear growth function $D(z)$~\citep{Marques20}. 

From the analyses of diverse cosmological tracers it is 
common to perform the measurements at scales of 
$R = 8$ Mpc$/h$, that is, $\sigma_{8,0} \equiv \sigma_{8}(z=0)$. %(see, e.g., \cite{Juszkiewicz10}). 
Thus, for the scale of $8$~Mpc$/h$ one has 
\begin{equation}\label{sig8}
\sigma_{8}(z) = D(z)\, \sigma_{8,0} \,.
\end{equation} 
Then, the product $[f \sigma_{8}](z)$ can be written as 
\begin{equation}
\label{fsig8}
[f \sigma_{8}](z) = -\,\sigma_{8,0}\, (1+z)\,\frac{dD(z)}{dz} \,,
\end{equation}
which directly measures the matter density perturbation rate. 

For the purpose of our analyses, one can obtain the function $\sigma_{8}(z)$ as the quotient of the functions 
\begin{equation}\label{eqn:sig8estimador}
\sigma_{8}^\text{q}(z)
\equiv \frac{[f \sigma_{8}]^{\mbox{\sc \,gp}}(z)}{f^{\mbox{\sc \,gp}}(z)} \,,
\end{equation}
where $f^{\mbox{\sc \,gp}}(z)$ and $[f \sigma_{8}]^{\mbox{\sc \,gp}}(z)$ were reconstructed using 
Gaussian Processes from measurements of $f(z)$ and $[f \sigma_{8}](z)$, respectively. 
The superscript `q' in $\sigma_{8}^\text{q}$
is used to indicate the quotient shown in equation (\ref{eqn:sig8estimador}). 

Once we obtain the function $\sigma_{8}^\text{q}(z)$, we shall obtain the function $S_8(z)$ through 
\begin{equation}\label{S8eq}
S_8(z) \equiv \sigma_{8}^\text{q}(z)  \left(\frac{\Omega_m(z)}{0.30}
\right)^{1/2} \,.
\end{equation}

%%---------------------------------
\section{Data set and Methodology}
\label{data}

In this section we present the $f(z)$ and $[f\sigma_{8}](z)$ data used to reconstruct first the $f^{\mbox{\sc \,gp}}(z)$ 
and $[f\sigma_{8}]^{\mbox{\sc \,gp}}(z)$ functions, then used to infer the cosmic evolution of the $\sigma_{8}(z)$ and $S_8(z)$ functions. 
In addition to these data, we use a set of $E(z)$ measurements performed by \cite{Ez2018a}, in the redshift interval 
$z\in[0.0,1.5]$, to reconstruct the $\gamma(z)$ function defined in equation (\ref{gammadef}).

\subsection{The $f(z)$ data}
%\section{Measurements of the growth rate function}

The literature reports diverse compilations of measurements of the growth rate of cosmic 
structures, $[f\sigma_{8}](z)$ \cite[see, e.g.][]{Basilakos12, Nunes16,Bessa21}, which we update here. 
Our compilation of $f(z)$ data, shown in table~\ref{table1}, follows these  criteria:
\\

(i) We consider $f(z)$ data obtained from uncorrelated redshift bins when the measurements concern the same cosmological tracer, and data from possibly 
correlated redshift bins when different cosmological tracers were analysed.
\\

(ii) We consider only data with a direct measurement of $f$, and not measurements of $f \sigma_8$ that use a fiducial cosmological model to eliminate the $\sigma_8$ dependence.
\\

(iii) We consider the latest measurement of $f$ when the same survey collaboration performed two or more measurements corresponding to diverse data releases. 

%%---------------------------------------------- table1 -----------------------------------------------------------
\begin{table*}\label{table-f}
	\centering
	\setlength{\extrarowheight}{0.11cm}
	\caption{Data compilation of 11 $f(z)$ measurements; see section~\ref{data} for details.
		%that shares important features, 
	}
	\label{table1}
	\begin{tabular}{|c|c|c|c|c|}
		\hline
		Survey &$z$ & $f$ & Reference & Cosmological tracer \\
		\hline
		ALFALFA & 0.013 & $0.56 \pm 0.07$ & \cite{Avila21} & HI extragalactic sources \\
		2dFGRS & 0.15 & $0.49 \pm 0.14$ & \cite{Hawkins03,Guzzo08} & galaxies \\
		GAMA & 0.18 & $0.49 \pm 0.12$ & \cite{Blake13} &  
		multiple-tracer: blue \& red gals. \\
		WiggleZ   & 0.22 & $0.60 \pm 0.10$ & \cite{Blake11} & galaxies \\
		SDSS    & 0.35 & $0.70 \pm 0.18$ & \cite{Tegmark06} & luminous red galaxies (LRG) \\
		GAMA   & 0.38 & $0.66 \pm 0.09$ & \cite{Blake13} & 
		multiple-tracer: blue \& red gals. \\
		WiggleZ & 0.41 & $0.70 \pm 0.07$ & \cite{Blake11} & galaxies \\
		2SLAQ & 0.55 & $0.75 \pm 0.18$ & \cite{Ross07} & LRG \& quasars \\
		WiggleZ & 0.60 & $0.73 \pm 0.07$ & \cite{Blake11} & galaxies \\
		VIMOS-VLT Deep Survey & 0.77 & $0.91 \pm 0.36$ & \cite{Guzzo08} & 
		faint galaxies  \\
		2QZ \& 2SLAQ & 1.40 & $0.90 \pm 0.24$ & \cite{DaAngela08} & quasars \\
		\hline
	\end{tabular}
\end{table*}
%%-----------------------------------------------table1-----------------------------------------------------------

\subsection{The $[f\sigma_{8}](z)$ data}

In table \ref{table2} we present our compilation of $f\sigma_{8}$ data. 
The criteria for selecting these data are:\\

(i) We consider $[f \sigma_8](z)$ data obtained from uncorrelated redshift bins when the measurements 
concern the same cosmological tracer, and data from possibly correlated redshift bins when different 
cosmological tracers were analysed. \\

(ii) We consider direct measurements of $f \sigma_8$. \\

(iii) We consider the latest measurement of $f \sigma_8$ when the same survey collaboration performed 
two or more measurements corresponding to diverse data releases. 

%%------------------------------------table2-------------------
%\linespread{1.25}
\begin{table*}\label{table-fsigma}
	\centering
	\setlength{\extrarowheight}{0.11cm}
	\caption{Data compilation of 20 $[f\sigma_8](z)$ measurements; see section~\ref{data} for details.
%(from Table II~\cite{Nesseris} + data from refs.~\cite{Aubert,Alam,Zhao,Nadathur}.
	}
	\label{table2}
	\begin{tabular}{|c|c|c|c|c|}
		\hline
		Survey & $z$ & $f \sigma_8$ & Reference & Cosmological tracer \\
		\hline
		SnIa+IRAS    & 0.02 & $0.398 \pm 0.065$ & \cite{Turnbull12} & SNIa + galaxies  \\
		6dFGS & 0.025  & $0.39 \pm 0.11$ & \cite{Achitouv17} & voids \\
		6dFGS & 0.067 & $0.423 \pm 0.055$ & \cite{Beutler12} & galaxies \\
		SDSS-veloc & 0.10  & $0.37 \pm 0.13$ & \cite{Feix15} & DR7 galaxies \\
		SDSS-IV & 0.15  & $0.53  \pm 0.16$ & \cite{Alam17} &$\!$eBOSS DR16 MGS \\
		%: $\!0.07 < z < 0.2$\\
		BOSS-LOWZ   & 0.32 & $0.384 \pm 0.095$ & \cite{Sanchez14} & DR10, DR11 \\
		SDSS-IV & 0.38  & $0.497 \pm 0.045$ & \cite{Alam17} & eBOSS DR16 galaxies \\
		%: $0.2 < z < 0.5$ \\
		WiggleZ   & 0.44  & $0.413 \pm 0.080$ & \cite{Blake12} & bright emission-line galaxies \\
		%$0.2 < z < 0.6$ \\
		CMASS-BOSS   & 0.57  & $0.453 \pm 0.022$ & \cite{Nadathur19} & DR12 voids+galaxies \\
		SDSS-CMASS & 0.59 & $0.488 \pm 0.060$ & \cite{Chuang16} & DR12 \\
%		VIPERS PDR-2 & 0.60 & $0.550 \pm 0.120$ & \cite{Pezzotta17} & galaxies \\
%$0.5 < z < 1.2$ \\
		SDSS-IV & 0.70  & $0.473 \pm 0.041$ &  \cite{Alam17} & eBOSS DR16 LRG \\
		WiggleZ  & 0.73  & $0.437 \pm 0.072$ & \cite{Blake12} & bright emission-line galaxies \\
		%$0.6 < z < 1.0$ \\
		SDSS-IV & 0.74  & $0.50 \pm 0.11$ &  \cite{Aubert20} & eBOSS DR16 voids \\
		VIPERS v7 & 0.76  & $0.440 \pm 0.040$ &  \cite{Wilson16} & galaxies \\
		SDSS-IV & 0.85  & $0.52 \pm 0.10$ &  \cite{Aubert20} & eBOSS DR16 voids \\
%		VIPERS PDR-2 & 0.86 & $0.400 \pm 0.110$ &  \cite{Pezzotta17} & galaxies \\
%$0.5 < z < 1.2$ \\
		SDSS-IV & 0.978 & $0.379 \pm 0.176$ &  \cite{Zhao19} & eBOSS DR14 quasars \\
		VIPERS v7 & 1.05  & $0.280 \pm 0.080$ &  \cite{Wilson16} & galaxies \\
		FastSound & 1.40 & $0.482 \pm 0.116$ &  \cite{Okumura16} & ELG \\
		SDSS-IV & 1.48  & $0.30 \pm 0.13$ &  \cite{Aubert20} & eBOSS DR16 voids \\
		SDSS-IV & 1.944 & $0.364 \pm 0.106$ &  \cite{Zhao19} & eBOSS DR14 quasars \\
		\hline
	\end{tabular}
	%\linespread{1.0}
\end{table*}
%%-----------------------------------------------table2-----------------------------------------------------------

\begin{figure*}
%\centering
\mbox{
\includegraphics[scale=0.6]{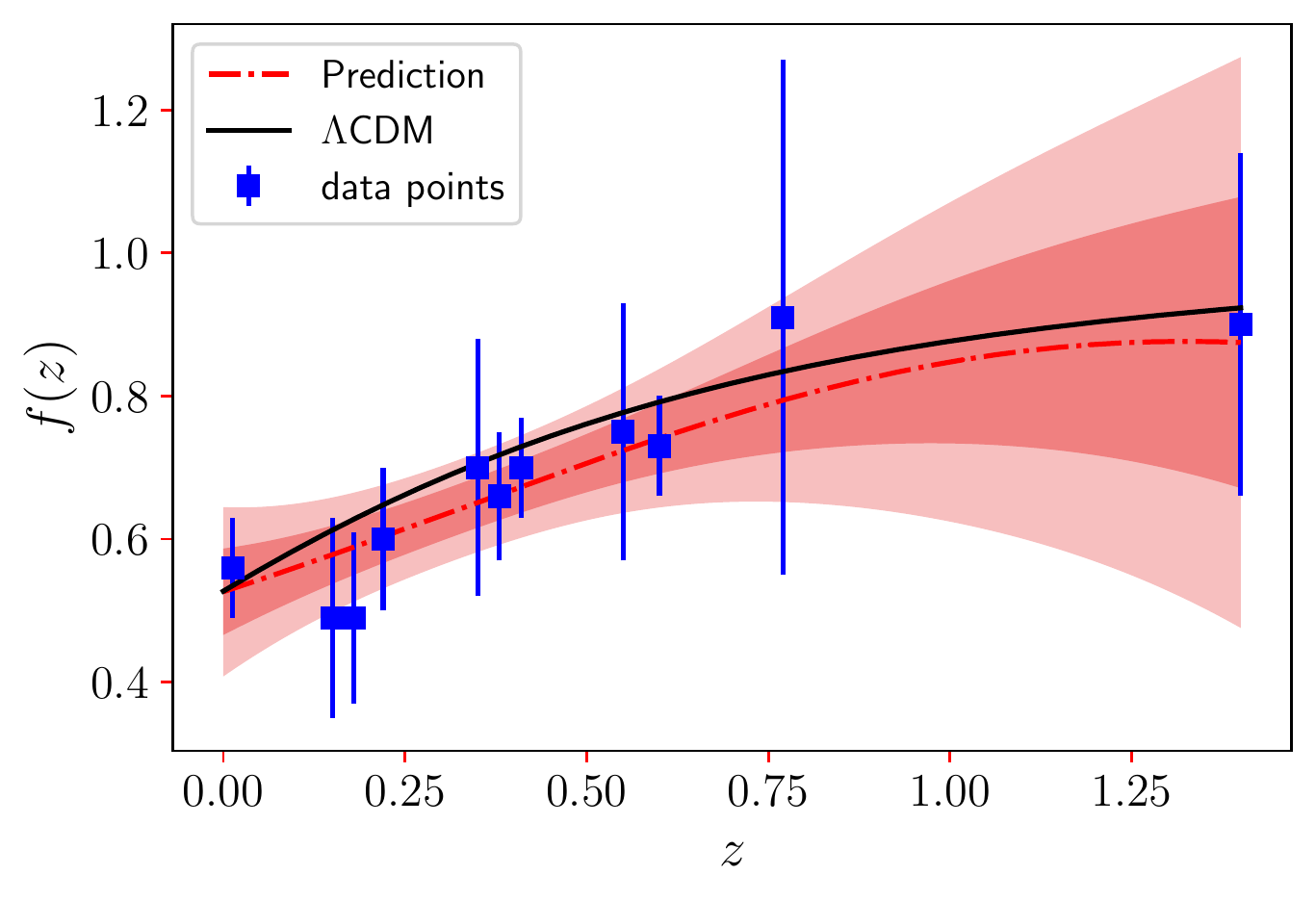} %\,\,\,\,\,\,
\hspace{0.2cm}
\includegraphics[scale=0.6]{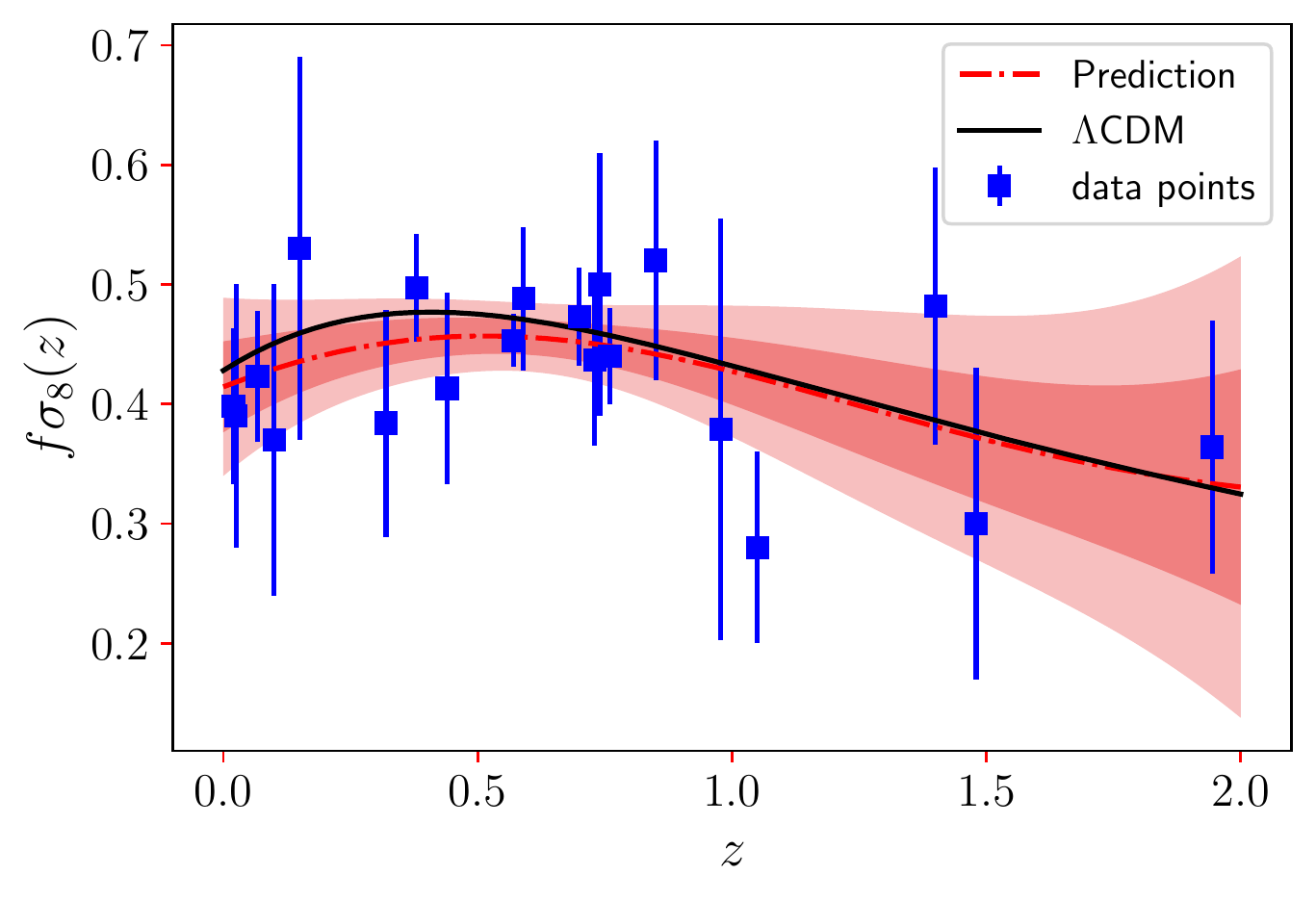}
}
\caption{Left panel: Growth rate reconstruction $f^{\mbox{\sc gp}}(z)$ from table \ref{table1}. 
Right panel: The reconstruction of $[f\sigma_{8}]^{\mbox{\sc gp}}(z)$ function using the sample in table \ref{table2}.
In both plots the shaded areas represent the $1\sigma$ and $2\sigma$ CL regions.}
	\label{fig:fzrecon}
\end{figure*}

%%-------------------------------------------------------------------
\subsection{Gaussian Processes Regression}

To extract maximum cosmological information from a given data set, as for instance the $f$ and $f\sigma_8$ data listed 
in Tables \ref{table1} and \ref{table2}, we perform a Gaussian Processes Regression (GP), obtaining in this way smooth 
curves for the functions $f^{\mbox{\sc gp}}(z)$ and $[f\sigma_8]^{\mbox{\sc gp}}(z)$ 
according to the approach described in section~\ref{sec2}. 
Both reconstructed functions are then used to obtain the cosmic evolution of $\sigma_{8}^\text{q}(z)$ and  $S_8(z)$. 

The GP consists of generic supervised learning method designed to solve regression and 
probabilistic classification problems, where we can interpolate the observations and 
compute empirical confidence intervals and a prediction in some region of 
interest~\citep{Rasmussen}. 
In the cosmological context, GP techniques has been used to reconstruct cosmological 
parameters, like the dark energy equation of state, $w(z)$, the expansion rate of the 
universe, the cosmic growth rate, and other cosmological functions 
(see, e.g.,~\cite{Seikel12,Shafieloo12,Javier16,Javier17, Zhang18,Marques19,Renzi20,Benisty20,Bonilla21a,Bonilla:2020wbn,Colgain21, Sun21,renzi2021resilience,bengaly2021null,Escamilla-Rivera2021,Dhawan2021,Mukherjee2021,Keeley2021,Huillier2020,Avila22,ruizzapatero2022modelindependent} 
for a short list of references). 

The main advantage in this procedure is that it is able to make a non-parametric inference using only 
a few physical considerations and minimal cosmological assumptions. Our aim is to reconstruct a function 
$F(x)$ from a set of its measured  values $F(x_i) \pm \sigma_i$, 
for different values $\{ x_i \}$ of the variable $x$. 
It assumes that the value of the function at any point $x_i$ follows a Gaussian distribution. 
The value of the function at $x_i$ is correlated with the value at other point $x_i'$. 
Thus, a GP is defined as
\begin{equation}
\label{eqn:GPs}
F(x_i)=\mathcal{GP}(\mu(x_i),\textrm{cov}[F(x_i),F(x_i)]) \,,
\end{equation}
where $\mu(x_i)$ and $\textrm{cov}[F(x_i),F(x_i)]$ are the mean and the variance of the variable at $x_i$, respectively. 
For the reconstruction of the function $F(x_i)$, the covariance between the values of this function at different positions 
$x_i$ can be modeled as
\begin{equation}
\label{eqn:cov}
\textrm{cov}[F(x),F(x')] = k(x,x') \,,
\end{equation}
where $k(x,x')$ is known as the kernel function. 
The kernel choice is often very crucial to obtain 
good results regarding the reconstruction of the function $F(x)$. 

The kernel most commonly used is the standard Gaussian Squared-Exponential (SE) approach, 
defined as
\begin{equation}
\label{eqn:kSE}
k_{\text{SE}}(x,x') = \sigma_F^2 \exp\left(-\frac{|x-x'|^2}{2 l^2}\right) \,,
\end{equation}
where $\sigma_{F}^2$ is the signal variance, which controls the strength of the correlation of the function $F$, 
and $l$ is the length scale that determines the capacity to model the main characteristics (global and local) 
of $F$ in the evaluation region ($l$ measures the coherence length of the correlation in $x$). 
These two parameters are often called hyper-parameters.

However, given the irregular pattern noticed in our data sets (observe the blue squares representing 
the $f(z)$ and $[f\sigma_8](z)$ data shown in the plots of figure \ref{fig:fzrecon}), 
a more general kernel is suitable for the GP analyses, namely the Rational Quadratic kernel (RQ), 
defined as \citep{Rasmussen}
\begin{equation}
\label{eqn:kRQ}
k_{\text{RQ}}(x,x') = \left(1+\frac{|x-x'|^2}{2\alpha l^2}\right)^{-\alpha} \,,
\end{equation}
where $\alpha$ is the scale mixture parameter. This kernel can be seen as an infinite sum of SE kernels with different characteristic length-scales.

Beside the choice of the kernel, the length scale bounds also have an influence in the results, as discussed in \cite{Sun21, Perenon21}. 
For data showing irregular pattern behavior, as the data we are considering for analyses, a more restrictive bounds for the 
hyper-parameters are necessary. 
To reconstruct the function $[f\sigma_8](z)$ correctly, our choice for the length scale 
bound corresponds to the redshift interval of the sample. 
For the $f\sigma_8$ sample, for instance, we fix the priors 
$0.1 \leq l \leq 2$ and $0.1 \leq \alpha \leq 2$.

It is worth mentioning that the choice of the kernel and the length scale parameters, $l$ and $\alpha$, were delicate steps for a robust GP reconstruction of the function $\gamma(z)$ from the $[f\sigma_8](z)$ data sample. 
However, the reconstructed functions $\sigma_8^\text{q}(z)$ and $S_8(z)$ were obtained robustly against those particular choices, 
and this is also true for the $\gamma(z)$ function reconstructed using the $f(z)$ and $E(z)$ data.

%%--------------------------------------------------------------
\section{Results and Discussions}
\label{results}

The left panel of figure~\ref{fig:fzrecon} shows the $f(z)$ reconstruction at $1 \sigma$ and $2 \sigma$ confidence levels (CL) 
in the redshift range $z\in[0.0, 1.4]$, and the blue squares are the data points from table \ref{table1}. 
The dash-dot line is the prediction obtained from the GP using the RQ kernel. 
When evaluated at the present time, we find $f(z=0) = 0.526 \pm 0.060$ at $1\sigma$ CL. 
In the right panel of  figure~\ref{fig:fzrecon} we quantify the same statistical information, but assuming our $[f\sigma_{8}](z)$ data sample. 
When evaluated at the present time, $z = 0$, we find $f\sigma_8(z=0) = 0.414 \pm 0.038$ at 1$\sigma$ CL. 
In both panels, the black solid line represents the $\Lambda$CDM prediction with the Planck-CMB best fit values~\citep{Aghanim:2018eyx}. 
One can notice that the model-independent obtained here from both data samples, tables~\ref{table1} and \ref{table2}, 
predicts a smaller amplitude in comparison with $\Lambda$CDM model, but globally compatible within $2 \sigma$ uncertainties.

\begin{figure*}
%\centering
\mbox{
\includegraphics[scale=0.6]{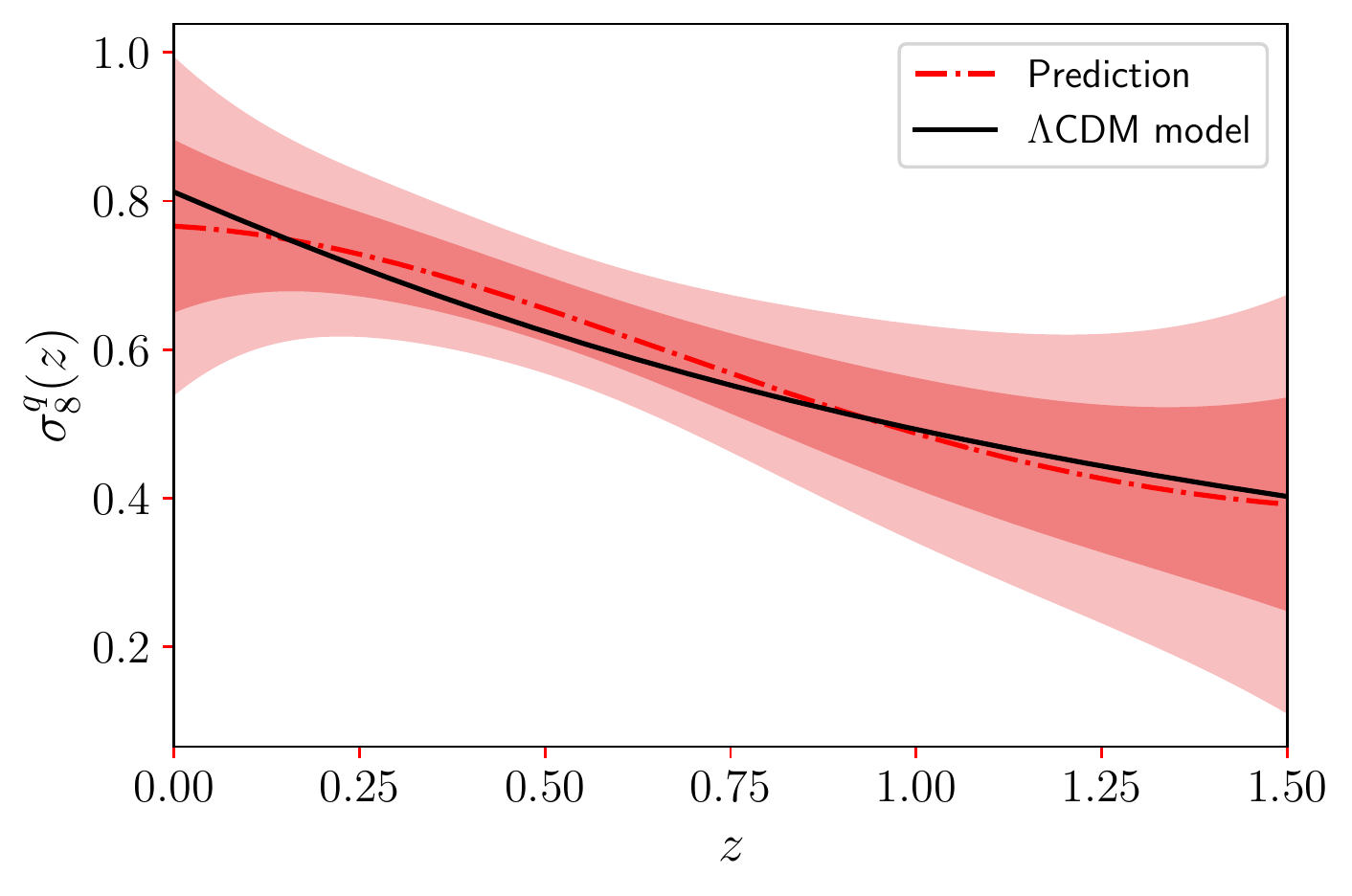} 
\hspace{0.cm}
\includegraphics[scale=0.6]{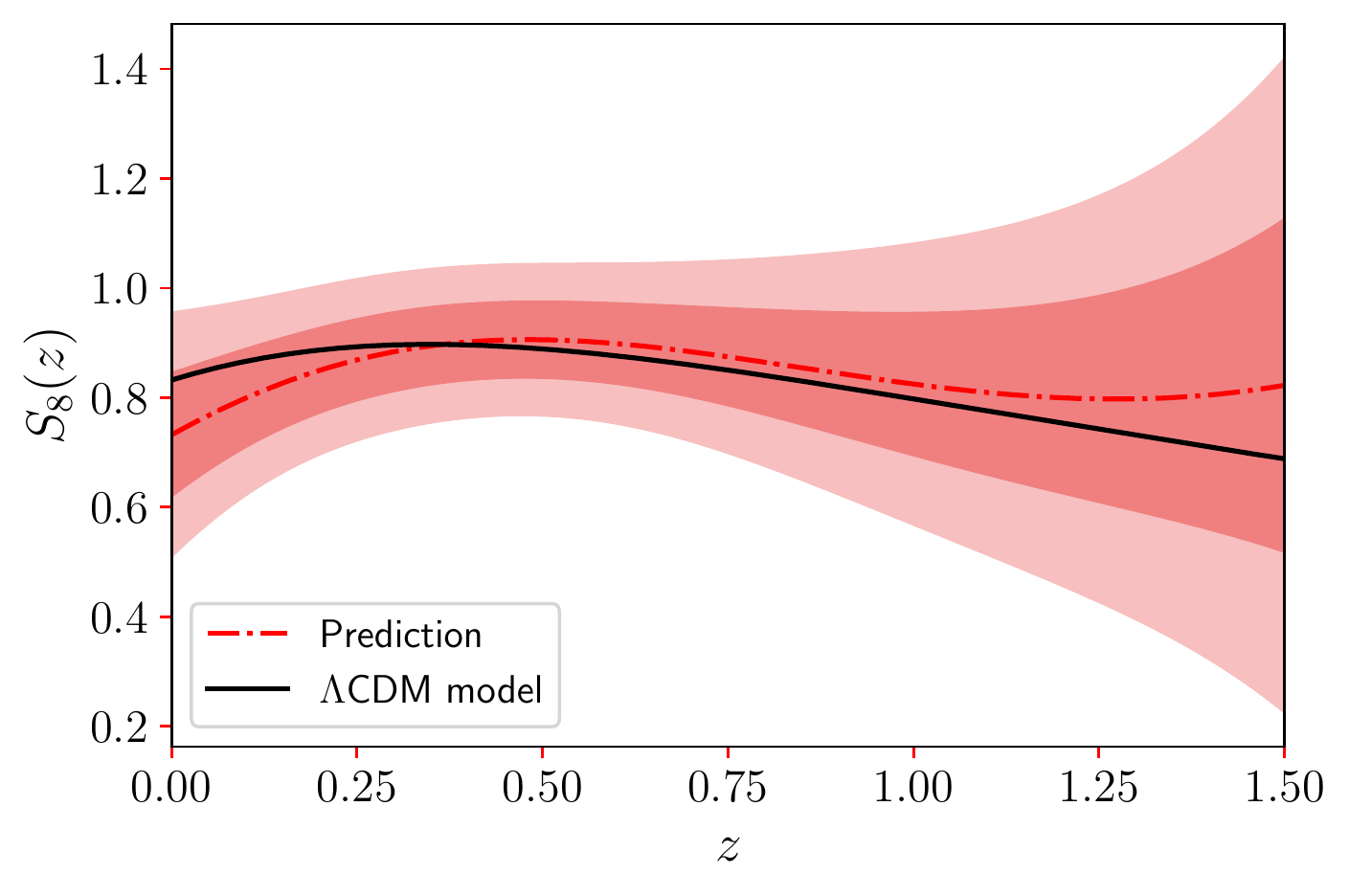}
}
\caption{Left panel: Reconstruction of the function $\sigma_{8}^\text{q}(z)$ (the letter $\text{q}$ to remember its origin: 
the quotient of two continuous functions) at $1 \sigma$ and $2 \sigma$ CL obtained from our 
$[f\sigma_{8}](z)$ and $f(z)$ data sample. 
The dot-dashed line represents the prediction from the data. 
Right panel: Same as in left panel, but for the $S_8(z)$ function. 
The black line represents the prediction from the $\Lambda$CDM model considering the Planck-CMB cosmological parameters.}
\label{fig:sig8recon}
\end{figure*}       

\begin{figure*}
%\centering
\mbox{
\includegraphics[scale=0.6]{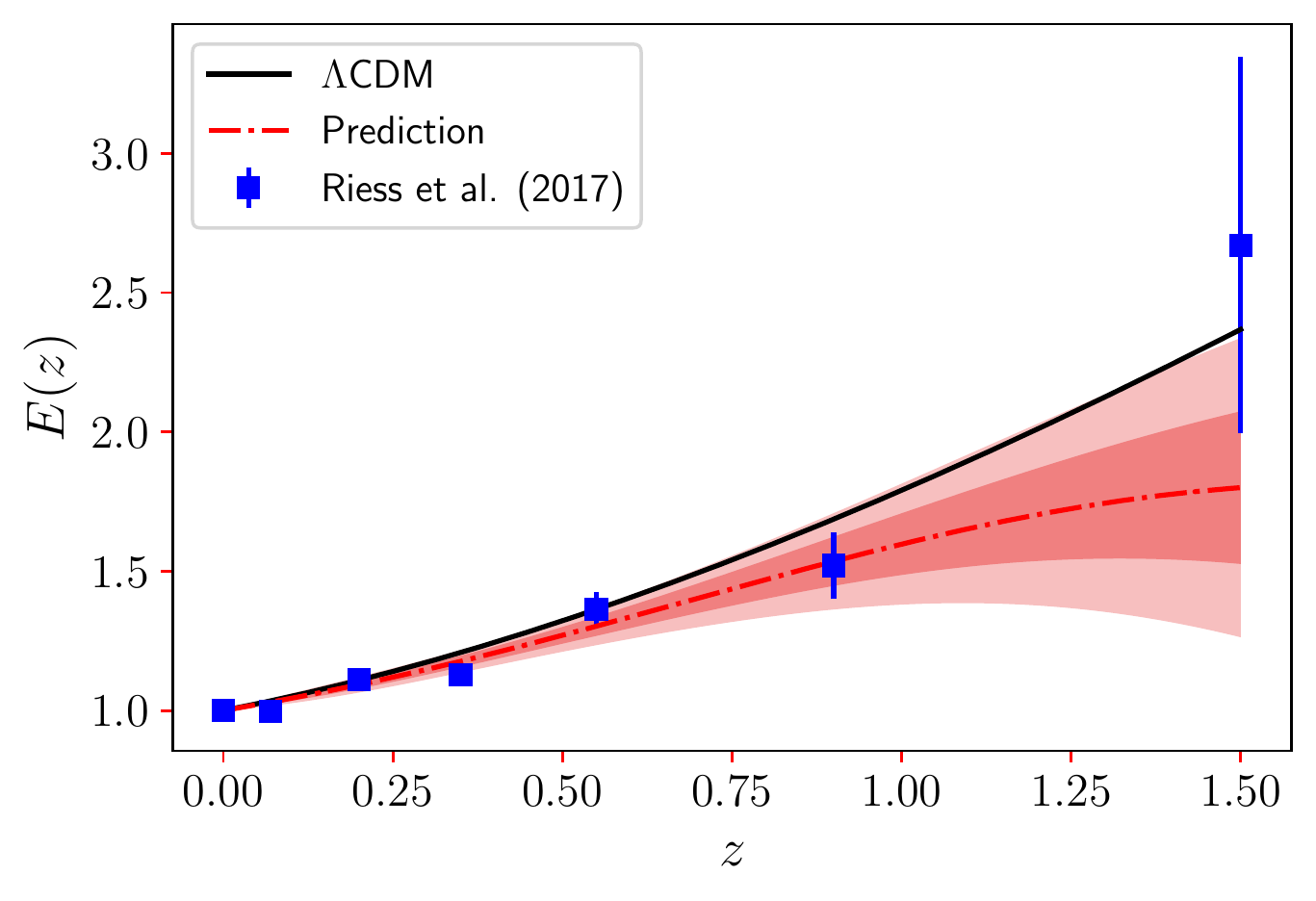} %\,\,\,\,\,\,\,\,
\hspace{0.cm}
\includegraphics[scale=0.6]{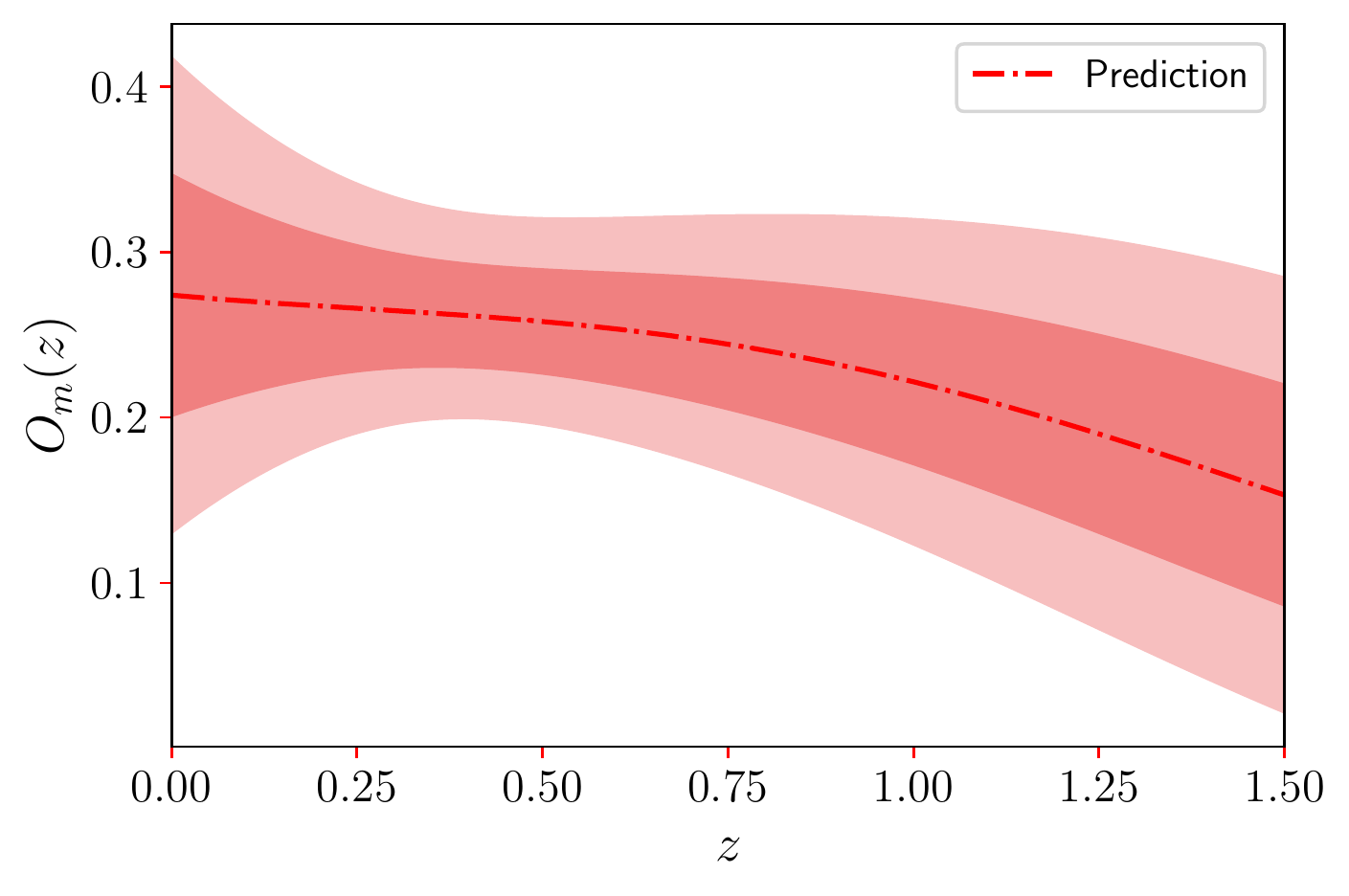}
}
\caption{Left panel: Reconstruction of the $E(z)$ function from the Pantheon sample. 
Right panel: Reconstruction of the $O_m(z)$ diagnostic test function from the Pantheon sample. 
The black line represents $\Lambda$CDM model from Planck-CMB values.}
\label{fig:ezrecon}
\end{figure*}

Figure~\ref{fig:sig8recon} on the left panel shows the function $\sigma_{8}^\text{q}(z)$ obtained through the methodology described in section 2. 
When evaluated at the present time, we find $\sigma_{8,0}^\text{q} = 0.766 \pm 0.116$ at  1$\sigma$ CL. 
On the right panel of figure~\ref{fig:sig8recon} we show the function $S_8$ obtained using $\sigma_{8}^\text{q}(z)$ according to equation (\ref{S8eq}). 
Here one notices that for such a  procedure we need infer also a reconstruction process for the function $\Omega_m(z)$. 
For this, in the context of the standard framework, we can use the $O_m(z)$ diagnostic function~\citep{Om2008}
\begin{equation}
\label{eqn:Omz}
O_m(z) = \frac{E^2(z) - 1}{(1 + z)^3 - 1} \,.
\end{equation}
If the expansion history $E(z)$ is driven by the standard $\Lambda$CDM model with null spatial curvature, then the function $O_m(z)$ 
is proportional to the matter density $\Omega_{m}(z)$. 
To reconstruct the $\Omega_{m}(z)$ function in minimal model assumptions, let us use the Supernovae Type Ia data from the Pantheon 
sample \citep{Scolnic:2017caz}. 
As is well known, the  Supernovae Type Ia traditionally have been one of the most important astrophysical tools in establishing the so-called standard cosmological model. For the present analyses, we use the Pantheon compilation, which consists of 1048 SNIa distributed
in the range $0.01 < z < 2.3$ \citep{Scolnic:2017caz}. 
With the hypothesis of a spatially flat Universe, the full sample of Pantheon can be binned into six model independent $E(z)$ data points \citep{Ez2018a}. 
We study the six data points reported by \cite{Ez2018b} in the form of $E(z)$, including theoretical and statistical considerations made by the
authors there for its implementation. 
Under these considerations, we find $\Omega_{m,0}=0.274 \pm 0.073$ at $1 \sigma$ CL. 
Note that this estimate is model-independent. 
Then, we reconstruct the evolution of the matter density in a model-independent way, by applying again the Pantheon sample on the definition 
$\Omega_m(z) \equiv \Omega_{m,0}(1+z)^{3}/E^2(z)$. 
Figure \ref{fig:ezrecon} on the left panel shows the robust reconstruction for the $E(z)$ function and on the right panel for the $O_m(z)$ 
diagnostic function. 
After these steps, we can infer the reconstruction for the $S_8$ function as a function of redshift (right panel in figure~\ref{fig:sig8recon}). 
When evaluated at the present time, we find $S_8(z=0) = 0.732 \pm 0.115$ at $1 \sigma$ CL. 

\begin{figure*}
%\centering
\mbox{
\includegraphics[scale=0.6]{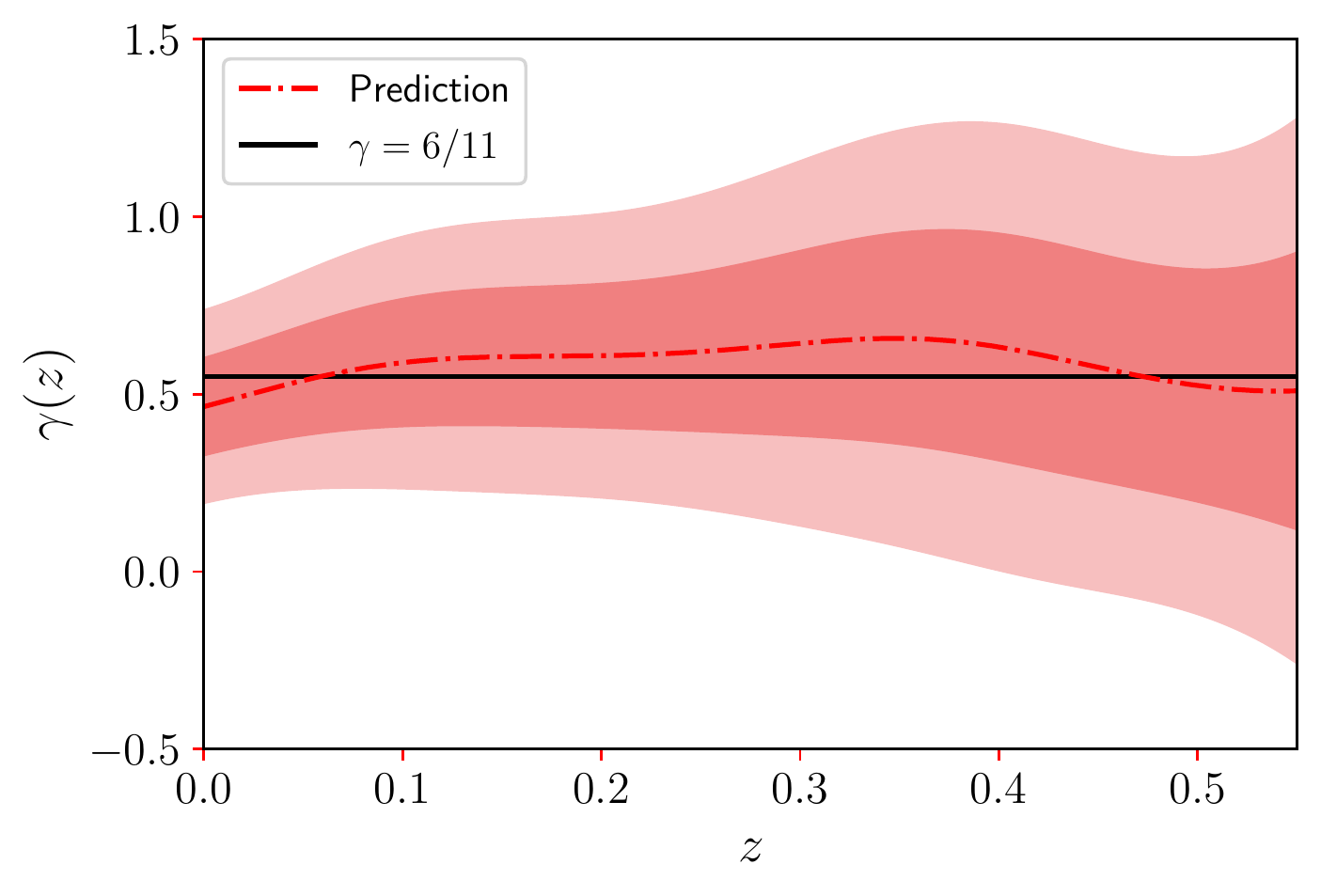} %\,\,\,\,\,\,\,
\hspace{0.cm}
\includegraphics[scale=0.6]{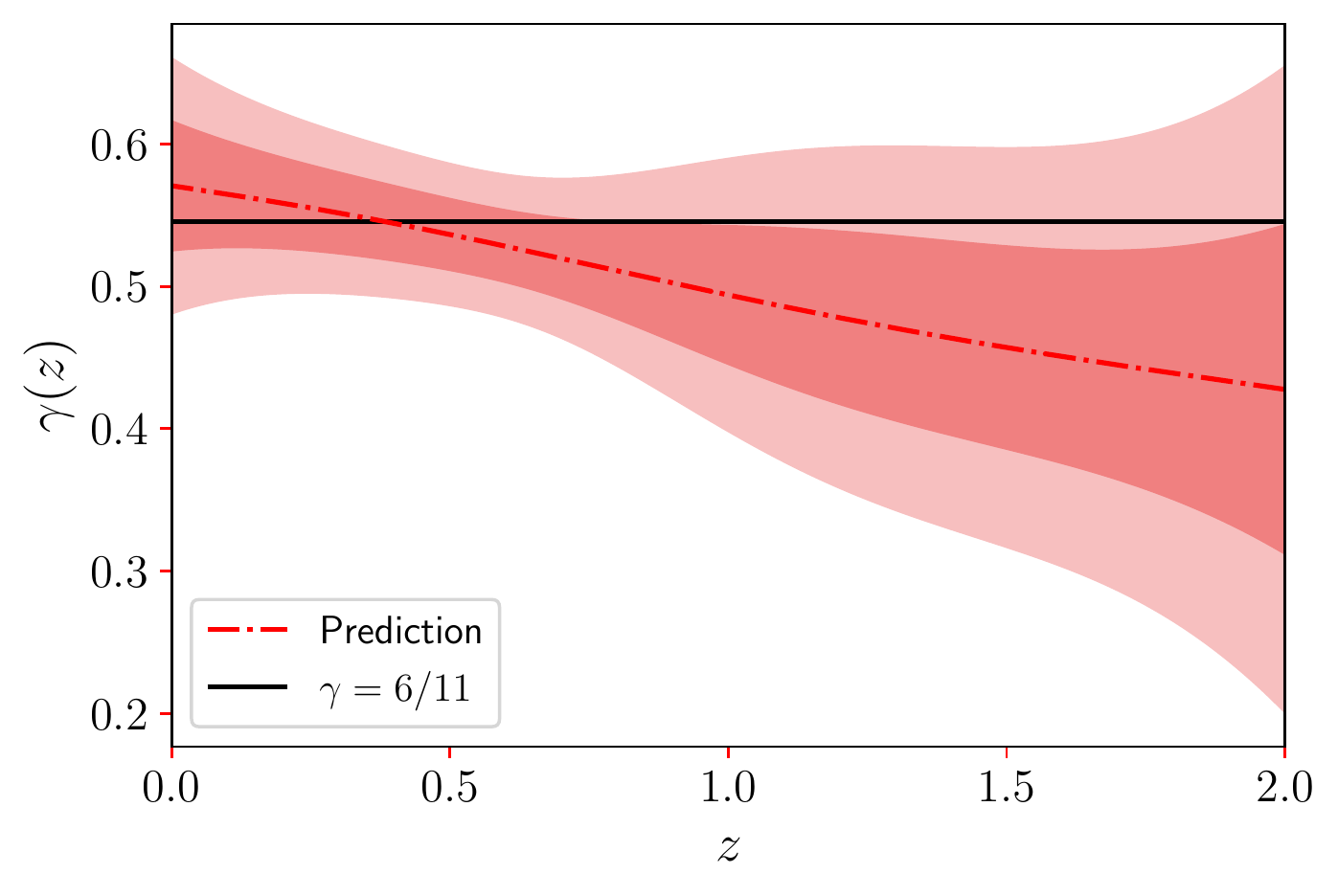}
}
\caption{Left panel: Reconstruction of the $\gamma(z)$ function with the $f(z)$ sample. 
Right panel: Reconstruction of the $\gamma(z)$ function with the $f\sigma_8$ sample.}
\label{fig:gammareconf}
\end{figure*}

\begin{figure*}
%\centering
\mbox{
\includegraphics[scale=0.85]{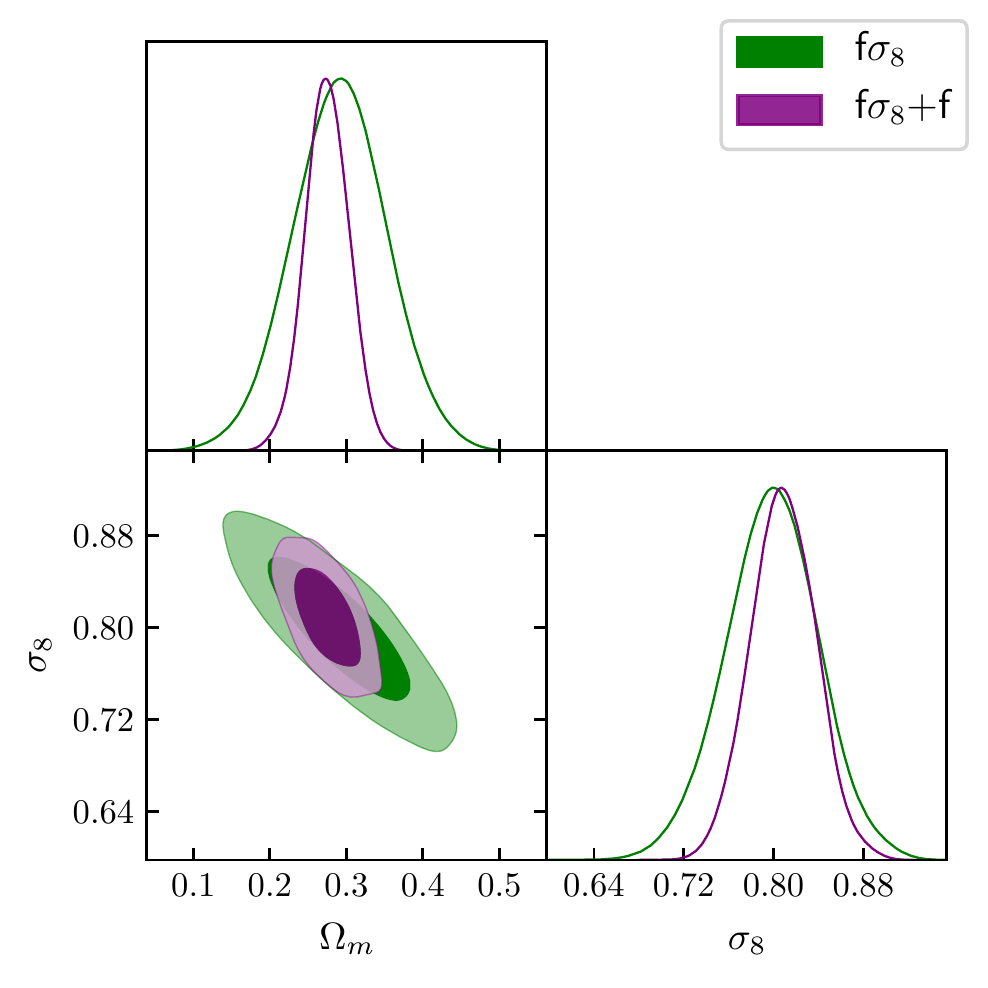} %\,\,\,\,\,\,\,
\hspace{0.cm}
\includegraphics[scale=0.85]{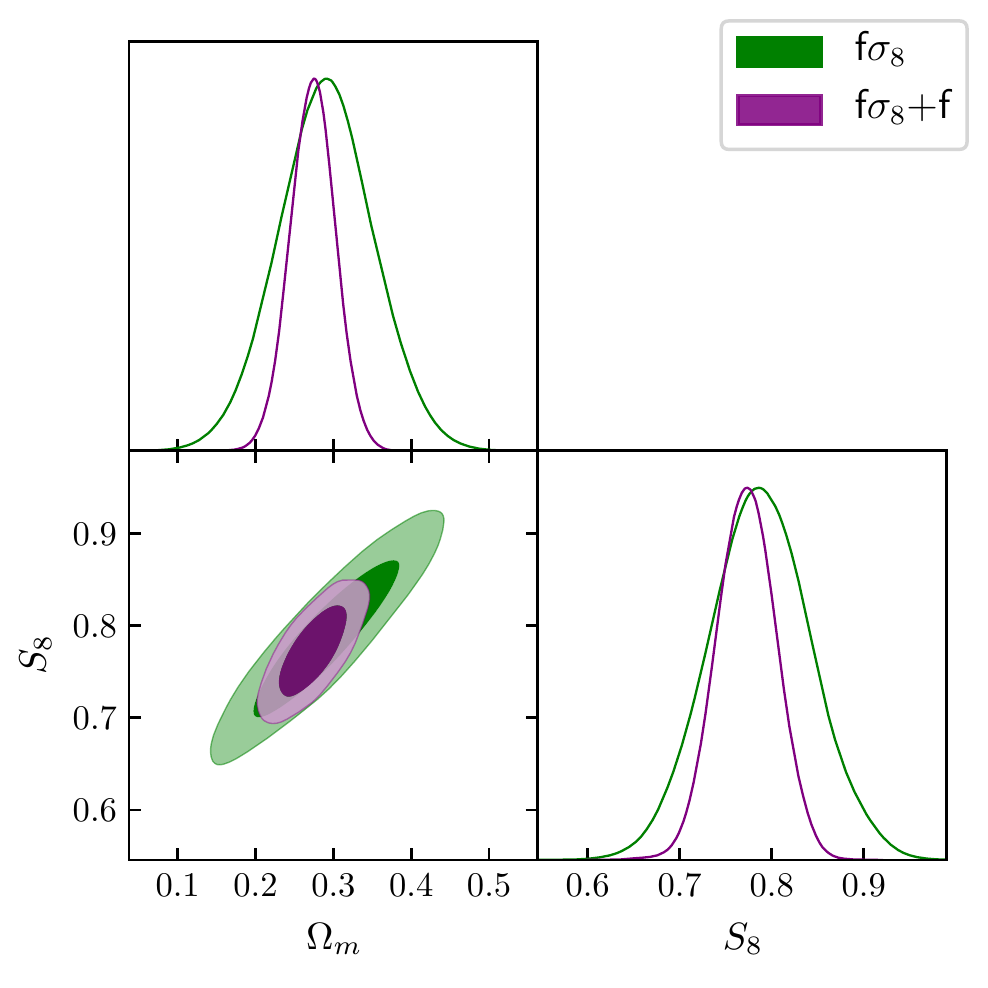}
}
\caption{1D and 2D posterior distributions for $[f\sigma_{8}](z)$ and $[f\sigma_{8}](z)$ + $f(z)$ data at $1 \sigma$ and $2 \sigma$ CL. 
Left panel: Contour diagram on the plane $\Omega_m - \sigma_8$. 
Right panel: Contour diagram on the plane $\Omega_m - S_8$.}
\label{fig:ParEsp}
\end{figure*}

Within the context of the $\Lambda$CDM model, CMB temperature fluctuations measurements from Planck and ACT+WMAP indicate $S_8$ values of $0.834 \pm 0.016$~\citep{Aghanim:2018eyx} and $0.840 \pm 0.030$~\citep{Aiola:2020azj}, respectively. On the other hand, the value of $S_8$ inferred by a host of weak lensing and galaxy clustering measurements is typically lower than the CMB-inferred values, ranging between $0.703$ to $0.782$: examples of surveys reporting lower values of $S_8$ include CFHTLenS~\citep{Joudaki:2016mvz}, KiDS-450~\citep{Joudaki:2016kym}, KiDS-450+2dFLenS~\citep{Joudaki:2017zdt}, KiDS+VIKING-450 (KV450)~\citep{Hildebrandt:2018yau}, DES-Y1~\citep{Troxel:2017xyo}, KV450+BOSS~\citep{Troster:2019ean}, KV450+DES-Y1~\citep{Joudaki:2019pmv,Asgari:2019fkq}, a re-analysis of the BOSS galaxy power spectrum~\citep{Ivanov:2019pdj}, KiDS-1000~\citep{Asgari:2020wuj}, and KiDS-1000+BOSS+2dFLenS~\citep{Heymans:2020ghw}. Planck Sunyaev-Zeldovich cluster counts also infer a rather low value of $S_8=0.774 \pm 0.034$~\citep{Ade:2015fva}. To balance the discussion, it is also worth remarking that KiDS-450+GAMA~\citep{vanUitert:2017ieu} and HSC SSP~\citep{Hamana:2019etx} indicate higher values of $S_8$, of $0.800^{+0.029}_{-0.027}$ and $0.804^{+0.032}_{-0.029}$, respectively. Also, combining data from CMB, RSD, X-ray, and SZ cluster counts,~\cite{Blanchard21} found $S_8 = 0.841\pm0.038$. From our overall results, summarized in figure~\ref{fig:sig8recon}, it can be noticed that our model-independent analyses are fully compatible with the Planck $\Lambda$CDM cosmology (prediction quantified by the black line in Figure~\ref{fig:sig8recon}). Because our approach does not assume any fiducial cosmology, the error bar estimate in $S_8$ is degenerate. Due to this, our model-independent estimates 
are also compatible with some weak lensing and galaxy clustering measurements.

Now, let us investigate the cosmic evolution of the growth index $\gamma(z)$. 
First, let us analyze and quantify its evolution as described by the definition given in equation (\ref{gammadef}). 
Figure \ref{fig:gammareconf} on the left panel shows $\gamma(z)$ at late times inferred from the $f(z)$ data in combination with 
the Pantheon sample. 
It is important to remember that the Pantheon sample is used to reconstruct the function $\Omega_m(z)$. 
The black line represents the prediction in GR theory. 
We find that $\gamma$ is still statistically compatible with GR. 
When evaluated at the present time, we find $\gamma(z=0) = 0.465 \pm 0.140$ at $1 \sigma$ CL.

On the other hand, following \cite{Arjona2020}, one can write the growth index $\gamma$ as a function of $[f \sigma_8]$ in the form

\begin{equation}
\gamma(a) = \frac{ \ln \left(\frac{[f \sigma_8](a)}{\int_0^{a} dx \frac{[f \sigma_8](x)}{x}} \right) }
{ \ln \left(\frac{a\,[f \sigma_8](a)^2}{3 \int_0^{a} dx [f \sigma_8](x) \int_0^x dy \frac{[f \sigma_8](y)}{y}} \right) } \,.
\end{equation}

The main advantage of the above equation is that it only requires $[f \sigma_8](a)$ data to describe $\gamma(a)$. 
In this way, we apply our data compilation, displayed in table \ref{table2}, in this equation and show our results in the right panel 
of figure \ref{fig:gammareconf}. 
When evaluated at the present time, we find $\gamma(z=0) =0.571 \pm 0.046$ at 1$\sigma$ CL. 
Note that both the data set and the  statistical approach developed here  are different from the analyses  presented in \cite{Arjona2020}. 
Although both reconstruction processes on $\gamma(z)$ are compatible with GR, it is interesting to note that data predictions show a different  tendency, while $f(z)$ data predict a behavior above the value $\gamma= 6/11$, for $z > 0.3$, the $[f\sigma_8](z)$ data sample predicts a behavior 
below $\gamma= 6/11$. 
Despite this, all analyses displayed  here are compatible with GR. That is, in short, we do not find any deviation from standard cosmology predictions.\\

It is worth commenting the  growth rate tension reported in the literature in light of recent statistical analyses, considering assumptions that could solve the Hubble and the growth rate tensions simultaneously.

A class of modified gravity theories that allows the Newton's  gravitational constant to evolve, i.e.  $G=G(z)$ evolves with $z$, 
can solve at the same time both the Hubble and growth rate tensions, as shown by~\citep{Perenon19,Marra21}. 
In \cite{Nesseris17}, 
parametrizing an evolving  gravitational constant, 
%using a parametrization $G=G(a)$, 
%$G_\text{eff}(a)$ 
the authors found no tension with the RSD data and  the Planck-$\Lambda$CDM model. 
Additionally, using an updated $f\sigma_8$ data set,~\cite{Kazantzidis18} shows that analysing a subsample of the 20 most recently published 
data the tension in $f\sigma_8$ disappears, and the GR theory is favoured over modified gravity theories.
%for the criterium selection of data  considering high-z data points %, i.e. $g_a=0$,

On the other hand, combining weak lensing, real space clustering and RSD data, \cite{Skara20} found a substantial increase in the growth tension: 
from $3.5\sigma$ considering only $f\sigma_8$ data to $6\sigma$ when taking into account also the $E_g$ data.

As a criterion for comparison, we look for previous studies in the growth rate tension using the GP reconstruction. 
In~\cite{Li21}, using $f\sigma_8$ data, the authors did not find any tension when no prior in $H_0$ is used in the analyses, which agrees with our results because no $H_0$ prior was assumed here. 
In~\cite{Alestas21}, the authors consider evolving dark energy 
models and show that, for these models, the growth rate tension
between dynamical probe data and CMB constraints increases. 
More recently \cite{Reyes22}, using different kernels for the GP reconstruction and two methodologies to obtain the hyperparameters, discovered  that the growth rate tension arises for specific redshift intervals and kernels.

Gaussian reconstruction is a powerful tool that allows to 
reconstruct functions from observational data without prior assumptions. 
However, it has the disadvantage that the reconstructed functions exhibit large uncertainties, 
as the case studied here where we have few data with large errors (see tables~\ref{table1} and~\ref{table2}). 
For example in~\cite{Quelle20}, using only a $f\sigma_8$ data set, the authors found no tension in the growth rate, but one observes that the confidence regions are large enough to  encompass different cosmological models. 
To avoid this inconvenience, the way adopted in the literature is to combine diverse cosmological probes or assume specific priors. 
From our results, and other statistical analyses like those  in~\cite{Li21} and \cite{Reyes22}, 
we can say that in the future, with more astronomical data measured with less uncertainty, the GP methodology may indeed solve the growth rate tension.

%%-------------------------------------------------------------------------------
\subsection{Consistency tests in $\Lambda$CDM}

It is important to perform consistency tests, comparing our results  with the predictions of the $\Lambda$CDM model. 
This time we search for $S_8$ and $\sigma_8$ but following a different approach. 
%
%It is interesting to compare the results of our analyses using a different approach. 
%%In order to compare our results with respect to the standard model $\Lambda$CDM, 
In fact, we now perform a Bayesian analysis with both data sets presented in the tables \ref{table1} and \ref{table2} 
using the Markov Chain Monte Carlo (MCMC) method to analyze the set of parameters $\theta_i = \lbrace \Omega_m,\sigma_8 \rbrace$, and 
building the posterior probability distribution function 
\begin{equation}
\label{L}
 p(D|\theta) \propto \exp \Big( - \frac{1}{2} \chi^2\Big) \, ,
\end{equation}
where $\chi^2$ is chi-squared function. The goal of any MCMC approach is to draw $M$ samples $\theta_i$ from the general posterior probability density
\begin{equation}
\label{psd}
p(\theta_i, \alpha|D) = \frac{1}{Z} p(\theta,\alpha) p(D|\theta,\alpha)  \, ,
\end{equation}
where $p(\theta,\alpha)$ and $p(D|\theta,\alpha)$ are the prior distribution and the likelihood function, respectively. Here, the quantities $D$ and $\alpha$ are the set of observations and possible nuisance parameters. 
The quantity $Z$ is a normalization factor. 
In order to constrain the baseline $\theta_i$, we assume a uniform prior such that: $\Omega_{m,0}$ $\in$ $\left[0.1,0.5 \right]$ and $\sigma_{8,0}$ $\in$ $\left[0.5,1.0 \right]$.

We perform the statistical analysis based on the \textit{emcee} \citep{Foreman_Mackey_2013} code along with \textit{GetDist} \citep{lewis2019getdist} to analyze our chains. We follow the Gelman-Rubin convergence criterion \citep{10.1214/ss/1177011136}, checking that all parameters in our chains had excellent convergence.

Figure \ref{fig:ParEsp} shows the posterior distribution in the parameter  space $\Omega_m - \sigma_8$ (Left panel) and $\Omega_m - S_8$ (Right panel) 
at $1 \sigma$ and $2 \sigma$ CL for $[f\sigma_{8}](z)$ and $[f\sigma_{8}](z)$ + $f(z)$ data set, respectively. 
For $\Lambda$CDM model, we find $\Omega_{m,0} = 0.292 \pm 0.061$, $\sigma_{8,0} = 0.798 \pm 0.040$ and $S_{8,0} = 0.788 \pm 0.055$ at $1 \sigma$ CL from $[f\sigma_{8}](z)$ only. 
When performing the joint analyses $[f\sigma_{8}](z)$ + $f(z)$, we find $\Omega_{m,0} = 0.274 \pm 0.029$, $\sigma_{8,0} = 0.809 \pm 0.029$ and 
$S_{8,0} = 0.773 \pm 0.033$ at $1 \sigma$ CL (for recent analyses see, e.g.~\cite{BonillaRivera:2016use,Nunes20b,Benisty21}). 

As well known, there is a tension for low-$z$ measurements of growth data, and it is weaker than the Planck-$\Lambda$CDM predictions (see \cite{Valentino_2021_S8,perivolaropoulos2021challenges} and reference therein for a review). 
Our results here also confirm that growth rate data based in our compilation and criteria also predict a suppression on the amplitude 
of the matter density perturbation at low $z$ due the low $\Omega_m$ estimation in comparison with that from the Planck-$\Lambda$CDM  baseline. 
Despite obtaining a low $\Omega_{m,0}$ best-fit value in our analyses, including the error estimates our results are in agreement with the Planck CMB 
cosmological parameters at $1 \sigma$ CL.

%%-------------------------------------------------------------------
\section{Final Remarks}
\label{sec5}

The study of the large-scale matter clustering in the universe is attracting interest of the scientific community due to valuable information encoded in the growth rate of cosmic structures, useful to discriminate between the standard model of cosmology and alternative scenarios. 
In this work we construct, using the GP algorithm, the cosmic evolution of the functions $\sigma_8(z)$, $S_8(z)$, 
and $\gamma(z)$ using sets of measurements of $f(z)$, $[f \sigma_8](z)$, and $E(z)$ 
(see tables~\ref{table1} and~\ref{table2}, and~\cite{Ez2018b}).

According to the current literature, measurements of the cosmological parameter $S_8(z=0)$ provided by early (using CMB) and late (through galaxy clustering at $z \lesssim 2$) cosmological tracers reveal some discrepancy between them, suggesting somehow that the process of cosmic structures growth could be different. Although this tension could be due to unknown --or uncalibrated-- systematics, it is worthwhile to investigate the possibility of new physics beyond the standard model.  This motivate us to construct the cosmic evolution of $\sigma_8^\text{q}(z)$ first, and then $S_8(z)$, using available data. 
All our results show a good concordance, at less than $2 \sigma$ CL, with the corresponding predictions derived from the standard cosmological model, 
i.e. the flat $\Lambda$CDM.

%To do this we use a supervised learning method, known as Gaussian Processes regression, to find firstly the continuous 
%functions $[f \sigma_8](z)$ and $f(z)$ and then divide both to get $\sigma_8(z) = [f \sigma_8](z)/f(z)$, as a function 
%of the redshift. 
%Our primary results, displayed in both panels of figure~\ref{fig:fzrecon}, shows the GP reconstructed continuous 
%functions $f(z)$ and $[f \sigma_8](z)$ from their corresponding data sets. 
%The next step is to obtain the continuous function $\sigma_8(z)$, shown in the left panel of figure~\ref{fig:sig8recon}, 
%as a quotient of $[f \sigma_8](z)$ over $f(z)$. 
%In the last step we find, using $\Omega_m(z)$ --inferred from $E(z)$ data-- and the reconstructed $\sigma_8(z)$, 
%the cosmic evolution of $S_8(z)$ up to $z \sim 1.5$ (see the right panel in figure~\ref{fig:sig8recon}). 
%Additionally, we also reconstruct the $\gamma(z)$ function in two cases: considering the $f(z)$ and the $[f \sigma_8](z)$ 
%data sets, respectively, as presented in the panels of figure~\ref{fig:gammareconf}. 

In the near future, we expect several percent measurements of the expansion history of the universe, as well as of the cosmic growth rate, in a large set of  experiments, e.g., through maps of the universe obtained by the Euclid satellite \citep{Amendola2013Euclid}, or  measuring the peculiar motions of galaxies using Type Ia supernovae from LSST \citep{Howlett2017}, RSD with DESI \citep{Hamaus2020}. 
Additionally, we will have the SKA telescopes performing BAO surveys and measuring weak gravitational lensing using 21 cm intensity mapping \citep{santos2015cosmology,bull2015measuring}. All of these efforts will either reveal a systematic cause or harden the current tension in the growth rate measurements. Then, the methodology and results presented here can be significantly improved with new and precise measurements. Therefore, we believe that future perspectives in obtaining estimates of $S_8$ minimally model-dependent with cosmic growth rate measurements can shed new light on the current $S_8$ tension. 
%%--------------------------------------------------------------------------------------------------------------

\begin{acknowledgements}
FA and AB thank CAPES and CNPq  for the grants under which this work was carried out. 
RCN acknowledges financial support from the Funda\c{c}\~{a}o de Amparo \`{a} Pesquisa 
do Estado de S\~{a}o Paulo (FAPESP, S\~{a}o Paulo Research Foundation) under the project no. 2018/18036-5.
\end{acknowledgements}

% BibTeX users please use one of

%\bibliographystyle{unsrt}
%\bibliographystyle{spphys}       % APS-like style for physics
\bibliographystyle{h-physrev}
\bibliography{example}   % name your BibTeX data base

\end{document}